\documentstyle[11pt] {amsart}

\begin{document}

\title[FORMS ON VECTOR BUNDLES AND THE CONFORMAL ANOMALY]
{FORMS ON VECTOR BUNDLES OVER HYPERBOLIC MANIFOLDS AND THE CONFORMAL ANOMALY}

\author{A. A. Bytsenko}
\address{Departamento de Fisica, Universidade Estadual de Londrina,
Caixa Postal 6001, Londrina-Parana, Brazil.\,\, {\em E-mail
address:} {\rm abyts@@uel.br}}

\author{E. Elizalde}
\address{Consejo Superior de Investigaciones Cient\'{\i}ficas\\
Institut d'Estudis Espacials de Catalunya (IEEC/CSIC)\\
Edifici Nexus, Gran Capit{\`a} 2-4, 08034 Barcelona, Spain; and
Departament ECM, Facultat de F\'{\i}sica\\
Universitat de Barcelona, Diagonal 647, 08028 Barcelona, Spain.
\,\, {\em E-mail address:} {\rm elizalde@@ieec.fcr.es}}

\author{R. A. Ulhoa}
\address{Departamento de Fisica, Universidade Estadual de Londrina,
Caixa Postal 6001, Londrina-Parana, Brazil}

\date{July 7, 2003}

\thanks{E.E. has been supported by DGI/SGPI (Spain), project
BFM2000-0810, and by CIRIT (Generalitat de Catalunya), contract
1999SGR-00257.}

\maketitle

\begin{abstract}

We study gauge theories based on abelian $p-$forms on real compact
hyperbolic manifolds. An explicit formula for the conformal
anomaly corresponding to skew--symmetric tensor fields is
obtained, by using zeta--function regularization  and the trace
tensor kernel formula. Explicit exact and  numerical values of the
anomaly for $p-$forms of  order up to $p=4$ in spaces of dimension
up to $n=10$ are then calculated.

\end{abstract}

\section{Introduction}

The conformal deformations of the Riemannian metric and the
corresponding conformal anomaly play an important role in quantum
theories. It is well known that evaluation of the conformal
anomaly is actually possible for even dimensional spaces albeit
its computation is extremely involved. The general structure of
such anomaly in curved even--dimensional spaces has been actively
studied (see, for example, Ref. \cite{Deser}). We briefly mention
here an analysis related to this phenomenon for constant curvature
spaces. The calculation of the conformal anomaly for the sphere
can be found in Ref. \cite{Copeland}. Explicit computations of the
anomaly (of the stress--energy tensor) for scalar and spinor
quantum fields in compact hyperbolic spaces have been carried out
in Refs. \cite{Bytsenko1,Bytsenko2} (see also Refs.
\cite{Elizalde,Bytsenko3}), using the zeta--function
regularization method \cite{zb2,zklaus,book}.

The purpose of this paper is to analyze the conformal anomaly
associated with tensor fields on real hyperbolic spaces. Skew
symmetric tensor fields play an important role in quantum field
theory, supergravity, and string theory, where they naturally
couple to two--form connections. Abelian two--forms are closely
related to the theory of gerbes, which plays a key role in string
theory \cite{Sharpe1}--\cite{Keurentjes}. Such forms can be
understood as a connection on an abelian gerbe. In the abelian
case (which will be considered in this paper) the self--dual
two--form can be easily reduced to the abelian one--form gauge
field. Generally, the covariant quantization of skew--symmetric
tensor fields has met difficulties with ghost counting and
BRST--transformations. In the framework of functional integration,
the covariant quantization of free generalized gauge fields,
$p-$forms and the BRST--transformations have been obtained in Ref.
\cite{Obukhov}.

In this paper we present a decomposition of the Hodge Laplacian
and the tensor kernel trace formula associated with free
generalized gauge fields ($p-$forms) on real hyperbolic spaces.
The main ingredient required is a type of differential form
structure on the physical, auxiliary, or ghost variables. We
consider spectral functions and the conformal anomaly associated
with physical degrees of freedom of the Hodge--de Rham operators
on $p-$forms.

\section{Quantum Dynamic of Exterior Forms in Hyperbolic Spaces}

We shall work with an $n-$dimensional compact real hyperbolic
space $X$ with universal covering $M$ and fundamental group
$\Gamma$. We can represent $M$ as the symmetric space $G/K$, where
$G=SO_1(n,1)$ and $K=SO(n)$ is a maximal compact subgroup of $G$.
Then we regard $\Gamma$ as a discrete subgroup of $G$ acting
isometrically on $M$, and we take $X$ to be the quotient space by
that action: $X=\Gamma\backslash M= \Gamma\backslash G/K$. Let
$\tau$ be an irreducible representation of $K$ on a complex vector
space $V_\tau$, and consider the induced homogeneous vector bundle
$G\times_K V_\tau$ (the fiber product of $G$ with $V_\tau$ over
$K$) $\longrightarrow M$ over $M$. Restricting the $G$ action to
$\Gamma$ we obtain the quotient bundle $E_\tau=\Gamma\backslash
(G\times_KV_\tau)\longrightarrow X=\Gamma\backslash M$ over $X$.
The natural Riemannian structure on $M$ (therefore on $X$) induced
by the Killing form $(\;,\;)$ of $G$ gives rise to a connection
Laplacian ${\frak L}$ on $E_\tau$. If $\Omega_K$ denotes the
Casimir operator of $K$, that is
\begin{equation}\label{01}
\Omega_K=-\sum y_j^2,
\end{equation}
for a basis $\{y_j\}$ of the Lie algebra ${\frak k}_0$ of $K$,
where $(y_j\;,y_\ell)=-\delta_{j\ell}$, then
$\tau(\Omega_K)=\lambda_\tau{\mathbf 1}$, for a suitable scalar
$\lambda_\tau$. Moreover, for the Casimir operator $\Omega$ of
$G$, with $\Omega$ operating on smooth sections $\Gamma^\infty
E_\tau$ of $E_\tau$, one has
\begin{equation}\label{02}
{\frak L}=\Omega-\lambda_\tau{\mathbf 1}\;
\end{equation}
(see Lemma 3.1 of \cite{Wallach}). For $\lambda\geq 0$, let
\begin{equation}\label{03}
\Gamma^\infty\left(X\;,E_\tau\right)_\lambda=
\left\{s\in\Gamma^\infty E_\tau\left|-{\frak L}s=\lambda s\right.
\right\}
\end{equation}
be the space of eigensections of ${\frak L}$ corresponding to
$\lambda$. Here we note that since $X$ is compact we can order the
spectrum of $-{\frak L}$ by taking $
0=\lambda_0<\lambda_1<\lambda_2<\cdots$, with
$\lim_{j\rightarrow\infty}\lambda_j=\infty$. We shall focus on the
more difficult (and interesting) case when $n=2k$ is even, and we
specialize $\tau$ to be the representation $\tau^{(p)}$ of
$K=SO(2k)$ on $\Lambda^p {\Bbb C}^{2k}$, say $p\neq k$. The case
when $n$ is odd will be dealt with later. It is convenient,
moreover, to work with the normalized Laplacian ${\frak
L}_p=-c(n){\frak L}$, where $c(n)=2(n-1)=2(2k-1)$. ${\frak L}_p$
has spectrum $\left\{c(n)\lambda_j\;,m_j\right\}_{j=0}^\infty$,
where the multiplicity $m_j$ of the eigenvalue $c(n)\lambda_j$ is
given by
\begin{equation}\label{04}
m_j={\rm
dim}\;\Gamma^\infty\left(X\;,E_{\tau^{(p)}}\right)_{\lambda_j}\;.
\end{equation}
Let $T_{j_1j_2...j_k}$ be a skew--symmetric tensor of
$(0,k)-$type, i.e.\,\,\,\,
$T_{\sigma(j_1,...,j_k)}\stackrel{def}{=}\,\,\, {\rm
sgn}(\sigma)T_{j_1,j_2,...,j_k}$, where ${\rm sgn}(\sigma)=\pm 1$
is the sign of the permutation $\sigma$. The exterior differential
$p-$form is
\begin{equation}
\omega_p = \frac{1}{p!}\sum_{j_1,...,j_p}T_{j_1j_2...j_p}
dx^{j_1}\wedge ... \wedge dx^{j_p} \mbox{.}
\end{equation}
Here $\wedge$ is the exterior product, $dx^j$ are the basis
one--forms, and $j= 1,2, ..., n$ (${\rm dim}\,M=n$). Let
$\Lambda^*(M)\equiv \oplus_{p=0}^{n}\Lambda^p$ be the graded
Cartan exterior algebra of differential forms, where $\Lambda^p$
is the space of all $p-$forms on $M$. Let $(*T)$ denote a skew
symmetric tensor of type $(0,n-k)$, i.e.
\begin{equation}
(*T)_{j_{k+1...j_n}} = \frac{1}{k!}\sqrt{|{\rm g}|}
\varepsilon_{j_1...j_n}T^{j_1... j_k}\,,\, \,\,\,\,\, T^{j_1...
j_k} ={\rm g}^{j_1\ell_1}\cdot\cdot\cdot {\rm
g}^{j_k\ell_k}T_{\ell_1... \ell_k} \mbox{,}
\end{equation}
where $\varepsilon_{j_1...j_n} = \pm 1$ for ${\rm sgn}(j_1...j_n)=
\pm 1$ is the Levi--Civita tensor density, and the metric ${\rm
g}_{j\ell}$ (external gravitational field) has the signature
$(+,+,...,+)$. In local coordinates the exterior differential, $d:
\Lambda^p \longrightarrow \Lambda^{p+1}$, and the
co--differential, $\delta: \Lambda^p\longrightarrow
\Lambda^{p-1}$, take respectively the forms
\begin{equation}
d\omega = \frac{1}{p!}\sum_{j_1,j_2,...,j_{p+1}} \frac{\partial
T_{j_2...j_{p+1}}}{\partial x^{j_1}} dx^{j_1}\wedge ... \wedge
dx^{j_{p+1}} \mbox{,}
\end{equation}
\begin{equation}
\delta\omega = - \frac{1}{(p-1)!}\sum_{j_1,j_2,...,j_{p-1}}
\frac{\partial T_{j_1...j_p}}{\partial x_{j_1}} dx^{j_2}\wedge ...
\wedge dx^{j_p} \mbox{.}
\end{equation}
\mbox{}From the last equations it is easy to prove the following
properties for operators and forms: $dd=\delta\delta=0$,\, $\delta
= (-1)^{np+n+1}*d*$,\, **$\omega_p = (-1)^{p(n-p)}\omega_p$. Let
$\alpha_p,\, \beta_p$ be $p-$forms; then, the invariant inner
product is defined by $(\alpha_p, \beta_p)\stackrel{def}{=}\int_M
\alpha_p\wedge*\beta_p$. The operators $d$ and $\delta$ are
adjoint to each other with respect to this inner product for
$p-$forms: $(\delta\alpha_p, \beta_p) = (\alpha_p, d\beta_p)$.
\\
The following result holds.

\medskip
\par \noindent
{\bf Theorem.}\,\,\, {\em For every $p=0,1,2,..., n$,
$\Lambda^n(M)$ admits the orthogonal direct sum decomposition
\begin{equation}
\Lambda^p(M) = d\Lambda^{p-1}(M)\oplus\delta
\Lambda^{p+1}(M)\oplus{\cal H}^p(M)
\mbox{,}
\end{equation}
where ${\cal H}^p$ is the space of all harmonic $p-$forms. That
means, every form $\omega_p \in \Lambda^p(M)$ can be written as
$\omega_p=\delta\omega_{p+1}+d\omega_{p-1}+h_p$ with $h_p$ being a
harmonic $p-$form, $\triangle h_p=0$. }
\\

This theorem is implied by the existence of the orthogonal sum
decomposition
\begin{equation}
\Lambda^p(M) = {\cal H}^p(M)\oplus \triangle \Lambda^p(M)
\mbox{.}
\end{equation}
It is known that the $L^2$ harmonic $p-$form $h_p^{(2)}$ appears
on even real hyperbolic manifolds only. In fact, the following
result holds: The manifold ${\Bbb H}^n$ admits $L^2$ harmonic
$p-$forms if and only if $n=2p$; for even dimensional real
hyperbolic manifolds the space of $L^2$ harmonic $p-$forms is
infinite dimensional (Ref. \cite{Donnelly}, p. 373). One can
consider the $L^2-$de Rham complex:
\begin{equation}
0\longrightarrow
\Lambda^0 (M)\stackrel{d_0}{\longrightarrow} ...
\longrightarrow \Lambda^p(M)\stackrel{d_p}{\longrightarrow}
\Lambda^{p+1}(M)\stackrel{d_{p+1}}{\longrightarrow} ...
\longrightarrow \Lambda^n(M) \longrightarrow 0
\mbox{,}
\end{equation}
and its associated $L^2-$cohomology
\begin{equation}
H^p(M)=\frac{{\rm ker}\,(\Lambda^p(M)\stackrel{d_p}{\longrightarrow}
\Lambda^{p+1}(M))}
{{\rm range}\,(d_{p-1}\Lambda^{p-1}(M))}
\mbox{.}
\end{equation}
A theorem of Kodaira (Ref. \cite{deRham}, p. 165) gives the
following injection:
\begin{equation}
h_p^{(2)}(M)\stackrel{j}{\longrightarrow} H^p(M)
\mbox{.}
\end{equation}
The map (injection) $j$ is an isomorphism if and only if $d_{p-1}$
has closed range. If $j$ is not an isomorphism, then $j$ has
infinite dimensional co--kernel. The associated Laplacian ${\frak
L}_p$ has closed range if and only if $d_p$ and $d_{p-1}$ have
closed range (Ref. \cite{Zucker}, p. 446).
\\
\\
In quantum field theory the Lagrangian associated with $\omega_p$
takes the form: $L=d\omega_p\wedge *d\omega_p$ (gauge field)\,,\,
$L=\delta\omega_p\wedge*\delta\omega_p$ (co--gauge field). The
Euler--Lagrange equations, supplied with the gauge, give: ${\frak
L}_p\omega_p =0\,,\,\,\delta\omega_p =0$ (Lorentz gauge);\,
${\frak L}_p\omega_p =0\,,\,\, d\omega_p =0$ (co--Lorentz gauge).
These Lagrangians give possible representation of tensor fields or
generalized abelian gauge fields. The two representations of
tensor fields are not completely independent, because of the
well--known duality property of exterior calculus which gives a
connection between star--conjugated gauge and co--gauge tensor
fields. The gauge $p-$forms are mapped into the co--gauge
$(n-p)-$forms under the action of the Hodge $*$ operator. The
vacuum--to--vacuum amplitude for the gauge $p-$form $\omega_p$
becomes \cite{Obukhov}:
\begin{equation}
Z = N\int D{\omega}\exp\left[
-(\omega, {\frak L}_p\omega)\right]
\prod_{j=1}^p
\left({\rm Vol}_{p-j}({\rm det}{\frak L}_{p-j})^{(j+1)/2}
\right)^{(-1)^{j+1}}
\mbox{,}
\end{equation}
where we need to factorize the divergent gauge group volume and
integrate over the classes of gauge transformations
($\omega \rightarrow \omega + d\phi$).

\section{The Trace Formula Applied to the Tensor Kernel}

Since $\Gamma$ is torsion free, each $\gamma\in\Gamma-\{1\}$ can
be represented uniquely as some power of a primitive element
$\rho:\gamma=\rho^{j(\gamma)}$ where $j(\gamma)\geq 1$ is an
integer and $\delta$ cannot be written as $\gamma_1^j$ for
$\gamma_1\in \Gamma$, \, with $j>1$ an integer. Taking
$\gamma\in\Gamma$, $\gamma\neq 1$, one can find $t_\gamma>0$ and
$m_{\gamma}\in {\frak M} \stackrel{def}{=}\{m_{\gamma}\in K |
m_{\gamma}a= am_{\gamma}, \forall a\in A\}$ such that $\gamma$ is
$G$ conjugate to $m_\gamma\exp(t_\gamma H_0)$, namely, for some
${\rm g}\in G$, \, one has ${\rm g}\gamma {\rm
g}^{-1}=m_\gamma\exp(t_\gamma H_0)$; that is, $\gamma$ is $G-$
conjugate to $m_\gamma\exp(t_\gamma H_0)$ and $m_\gamma\in
SO(n-1)$. For $\mbox{Ad}$ denoting the adjoint representation of
$G$ on its complexified Lie algebra, one can compute $t_\gamma$ as
follows \cite{Wallach1}:
\begin{equation}\label{09b}
e^{t_\gamma}={\rm max}\left\{
|c|\left|c= {\rm an\,\,\, eigenvalue\,\,\, of\,\,\,\, Ad}
(\gamma):{\rm g}\rightarrow {\rm g}\right.
\right\}\;.
\end{equation}
Also, $\gamma=\delta^{j(r)}$, where $j(r)\geq 1$ is a whole number
and $\delta\in\Gamma-\{1\}$ is a primitive element; ie. $\delta$
can not be expressed as $\gamma_1^j$ for some $\gamma_1\in\Gamma$
and some integer $j>1$. The pair $(j(\gamma),\;\delta)$ is
uniquely determined by $\gamma\in \Gamma-\{1\}$. These facts are
known to follow since $\Gamma$ is torsion free.

Let $a_0, n_0$
denote the Lie algebras of $A, N$ in an Iwasawa decomposition $G=KAN$.
The complexified Lie algebra ${\rm g}={\rm g}_0^{\Bbb C}
={so}(2k+1, {\Bbb C})$ of $G$ is of Cartan type $B_k$ with the Dynkin diagram
\begin{equation}
  \underbrace{\bigcirc-\bigcirc-\bigcirc \cdots \bigcirc-\bigcirc}_{2k\,
   {\rm  nodes}} = \bigcirc\,\,.
\end{equation}
\\
Since the rank of $G$ is one,
$\dim a_0=1$ by definition, say $a_0={\Bbb R}H_0$ for a suitable basis
vector $H_0$:
\begin{equation}
H_0 = \left[ \begin{array}{ll}
0\,\,\,\,\, 0\,\,\,\,\,\,\,\,\:\: .\,\,.\,\,.\,\,\,\,\,\,\,\,\:\: 0\, &1\\
0                                                &0\\
.                                                &.\\
.                                                &.\\
.                                                &.\\
0                                                &0\\
1\,\,\,\,\, 0\,\,\,\,\,\,\,\,\:\: .\,\,.\,\,.\,\,\,\,\,\,\,\,\:\: 0\, &0
\end{array} \right]
\mbox{,}
\end{equation}
\\
is a $(k+1)\times(k+1)$ matrix. With this choice we have the
normalization $\beta(H_0)=1$, where $\beta: a_0\rightarrow{\Bbb
R}$ is the positive root which defines $n_0$ (for more details see
Ref. 15). Define now $C(\gamma)$ on $\Gamma-\{1\}$ by
\begin{equation}
C(\gamma)\stackrel{def}=e^{-\rho_0t_\gamma}|\mbox{det}_{n_0}\left(\mbox{Ad}
(m_\gamma
e^{t_\gamma H_0})^{-1}-1\right)|^{-1}\mbox{.}
\end{equation}
Finally, let $C_\Gamma\subset\Gamma$ be a complet set of
representations in $\Gamma$ of its conjugacy classes. This means
that any two elements in $C_\Gamma$ are non--conjugate, and any
$\gamma\in \Gamma$ is $\Gamma-$conjugate to some element
$\gamma_1\in C_\Gamma:x\gamma x^{-1}=\gamma_1$ for some $x\in
\Gamma$. The reader may consult the appendix of \cite{Williams}
for further structural data concerning the Lie group $SO_1(n,1)$
(and other rank one groups).
\\
\\
Let $\tau = \tau_k=$ representation of $K$ on $\Lambda^j{\Bbb
C}^{2k}$. The space of smooth sections $\Gamma^\infty E_\tau$ of
$E_\tau$ is just the space of smooth $p-$forms on $X$. We can
therefore apply the version of the trace formula developed by
Fried in Ref. \cite{Fried}. First we set up some additional
notation. For $\sigma_j$ the natural representation of $SO(2k-1)$
on $\Lambda^j {\Bbb C}^{2k-1}$, one has the corresponding
Harish--Chandra--Plancherel density given ---for a suitable
normalization of the Haar measure $dx$ on $G$--- by
\begin{equation}\label{07}
\mu_{\sigma_p(r)}=
\frac{\pi}{2^{4k-4}[\Gamma(k)]^2}
\left(
\begin{array}{c}
2k-1\\ p
\end{array}
\right)
rP_{\sigma_p}(r)\tanh(\pi r)\;,
\end{equation}
\\
for $0\le p \le k-1$, where
\begin{equation}\label{08}
P_{\sigma_p}(r)=\prod_{\ell=2}^{p+1}
\left[
r^2+\left(k-\ell+\frac{3}{2}\right)^2
\right]
\prod_{\ell=p+2}^{k}
\left[
r^2+\left(k-\ell+\frac{1}{2}\right)^2
\right]\;
\end{equation}
\\
is an even polynomial of degree $2k-2$. One has that
$P_{\sigma_p}(r)= P_{\sigma_{2k-1-p}}(r)$ and
$\mu_{\sigma_p}(r)=\mu_{\sigma_{2k-1-p}}(r)$ for $k\le p\le 2k-1$.
Define the Miatello coefficients \cite{Miatello,BytsenkoX}
$a_{2\ell}^{(p)}$ for $G=SO_1(2k+1, 1)$ by $
P_{\sigma_p}(r)=\sum_{\ell=0}^{k-1}a_{2\ell}^{(p)}r^{2\ell}\;,
0\le p\le 2k-1 $.
\\
\\

Let ${\rm Vol}(\Gamma\backslash G)$ denote the integral of the
constant function $\mathbf{1}$ on $\Gamma\backslash G$ with
respect to the $G-$invariant measure on $\Gamma\backslash G$
induced by $dx$. For $0\leq p\leq n-1$ the Fried trace formula
applied to kernel ${\mathcal K}_t$ holds \cite{Fried}:
\begin{equation}
{\rm Tr}\left(e^{-t{\frak L}_{p}}\right)=I_{\Gamma}^{(p)}({\mathcal
K}_t)
+I_{\Gamma}^{(p-1)}({\mathcal K}_t)
+H_{\Gamma}^{(p)}({\mathcal K}_t)+
H_{\Gamma}^{(p-1)}({\mathcal K}_t)
\mbox{,}
\end{equation}
where $I_{\Gamma}^{(p)}({\mathcal K}_t)\,,
H_{\Gamma}^{(p)}({\mathcal K}_t)$ are the identity and hyperbolic
orbital integrals, respectively,
\begin{equation}
I_{\Gamma}^{(p)}({\mathcal K}_t)
\stackrel{def}{=}\frac{\chi(1){\rm Vol}
(\Gamma\backslash G)}{4\pi}
\int_{\Bbb R}dr\,\mu_{\sigma_p}(r)e^{-t(r^2+p+\rho_0^2)}
\mbox{,}
\end{equation}
\begin{equation}
H_{\Gamma}^{(p)}({\mathcal K}_t)
\stackrel{def}{=}\frac{1}{\sqrt{4\pi t}}
\sum_{\gamma\in C_
\Gamma-\{1\}}\frac{\chi(\gamma)}{j(\gamma)}t_\gamma C(\gamma)
\chi_{\sigma_p}
(m_\gamma)
\exp\left\{-t(\rho_0^2+p)-
t_\gamma^2/(4t)\right\}
\mbox{,}
\end{equation}
with $\rho_0=(n-1)/2$, and $\chi_\sigma(m)={\rm Tr}\;\sigma(m)$
for $m\in SO(2n-1)$.
\\
For $p\geq 1$ there is a measure $\mu_{\sigma}(r)$ corresponding
to a general irreducible representation $\sigma$ of ${\frak M}$.
Let $\sigma_p$ be the standard representation of ${\frak
M}=SO(n-1)$ on $\Lambda^p{\Bbb C}^{(n-1)}$. If $n=2k$ is even then
$\sigma_p\,\,(0\leq p\leq n-1)$ is always irreducible; if $n=2k+1$
then every $\sigma_p$ is irreducible except for $p=(n-1)/2=k$, in
which case $\sigma_k$ is the direct sum of two spin--$(1/2)$
representations
$\sigma^{\pm}:\,\,\sigma_k=\sigma^{+}\oplus\sigma^{-}$. For $p=k$ the
representation $\tau_k$ of $K=SO(2k)$ on $\Lambda^k {\Bbb C}^{2k}$
is not irreducible: $\tau_k=\tau_k^{+}\oplus\tau_k^{-}$ is the
direct sum of two spin--$(1/2)$ representations.
\\
\\

\noindent{\bf The case of the trivial representation}. In the case
of the trivial representation ($p=0$, i.e. for smooth functions or
smooth vector bundle sections) the measure $\mu(r)\equiv
\mu_{0}(r)$ corresponds to the trivial representation of ${\frak
M}$. Therefore, we take $I_{\Gamma}^{(-1)}({\mathcal K}_t)
=H_{\Gamma}^{(-1)}({\mathcal K}_t)=0$. Let $\chi_{\sigma}(m) =
{\rm trace}(\sigma(m))$ be the character of $\sigma$, for $\sigma$
a finite--dimensional representation of ${\frak M}$. Since
$\sigma_0$ is the trivial representation, one has
$\chi_{\sigma_0}(m_{\gamma})=1$. In this case, formula (15)
reduces exactly to the trace formula for $p=0$
\cite{Wallach,Elizalde,Bytsenko3,Williams,Bytsenko4},
\begin{equation}
I_{\Gamma}^{(0)}({\mathcal K}_t)
=\frac{\chi(1)\mbox{vol}(\Gamma\backslash G)}
{4 \pi}\int_{\Bbb R}dr\,\mu_{\sigma_0}(r)e^{-t(r^2+\rho_0^2)}d
\end{equation}
The function
$H_{\Gamma}^{(0)}({\mathcal K}_t)$ has the form
\begin{equation}
H_{\Gamma}^{(0)}({\mathcal K}_t)
=\frac{1}{\sqrt{4\pi t}}
\sum_{\gamma\in C_
\Gamma-\{1\}}\chi(\gamma)t_\gamma j(\gamma)^{-1}
C(\gamma)\exp \{-t\rho_0^2-t_\gamma^2/(4t)\}
\mbox{.}
\end{equation}

\section{Spectral Functions on $p-$Forms and the Conformal Anomaly}

The transverse part of the skew symmetric tensor is represented by
the co--exact $p-$form $\omega_p^{(CE)}=\delta\omega_{p+1}$, which
trivially satisfies $\delta\omega_p^{(CE)}=0$, and we denote by
${\frak L}_p^{(CE)} =\delta d$ the restriction of the Laplacian on
the co--exact $p-$form. The goal now is to extract the co--exact
$p-$form on the manifold which describes the physical degrees of
freedom of the system. Choosing a basis $\{\omega_p^{\ell}\}$ of
$p-$forms (eigenfunctions of the Laplacian), we get
\cite{Bytsenko1a,Bytsenko5,Bytsenko1b}
\begin{eqnarray}
{\rm Tr}\left(e^{-t{\frak L}_{p}^{(CE)}}\right)
&=&
\sum_{j=0}^p(-1)^j \left(
I_{\Gamma}^{(p-j)}({\mathcal K}_t)
+I_{\Gamma}^{(p-j-1)}({\mathcal K}_t) \right. \notag \\
&+& \left. H_{\Gamma}^{(p-j)}({\mathcal K}_t)+
H_{\Gamma}^{(p-j-1)}({\mathcal K}_t)-b_{p-j} \right)
\mbox{,}
\end{eqnarray}
where $b_j$ are the Betti numbers,
$b_j \equiv b_j(M) = {\rm rank}_{{\Bbb Z}}H_j(M; {\Bbb Z})$.
\\

For constant conformal deformations of the
Riemannian metric ${\rm g}^{\mu\nu}$ the variation
of the connected vacuum functional ${\frak W}$ can be expressed in terms of the
generalized zeta function $\zeta(s|{\frak A})$ \cite{Hawking,Dowker} associated
with the Laplace--Beltrami type operator ${\frak A}$,
\begin{equation}
\delta {\frak W} = - \zeta (0|{\frak A}) {\rm log} \mu^2 = (1/2)\int
d({\frak Vol})\,< T_{\mu\nu}(x)>\delta {\rm g}^{\mu \nu} (x)
\mbox{,}
\label{tensor}
\end{equation}
where $\mu$ is a
renormalization mass parameter and $<T_{\mu\nu} (x)>$ means that all
connected vacuum graphs of the stress--energy tensor $T_{\mu\nu} (x)$
are to be included. Then Eq. (\ref{tensor}) leads to the result
\begin{equation}
<T_\mu^\mu (x) > =  {\frak Vol}^{-1} \zeta (0|{\frak A}),
\label{tensor1}
\end{equation}
where for ${\Bbb S}^n$:\, ${\frak Vol} = 2\pi^{(n+1)/2}\, R^n
/(\Gamma ((n+1)/2)$, while for the compact manifold
$\Gamma\backslash {\Bbb H}^n$\,: \, ${\frak Vol} = {\rm
Vol}(\Gamma\backslash G)\, R^n$, where $R$ is the radius
corresponding to the compact space.

Our goal now is to calculate the value of generalized zeta function

$$
\!\!\!\!\!\!\!\!\!\!\!\!\!\!\!\!\!\!\!\!\!\!\!\!\!\!\!
\zeta(s|{\frak L}_{p}^{(CE)}) = \frac{1}{\Gamma(s)}
\int_0^{\infty}dt\,t^{s-1} {\rm Tr}\,e^{-t{\frak L}_{p}^{(CE)}} =
\sum_{j=0}^p\frac{(-1)^J}{\Gamma(s)}\int_0^{\infty}dt\,t^{s-1}
$$

\begin{equation}
\times
\left(
I_{\Gamma}^{(p-j)}({\mathcal K}_t)
+I_{\Gamma}^{(p-j-1)}({\mathcal K}_t)
+H_{\Gamma}^{(p-j)}({\mathcal K}_t)+
H_{\Gamma}^{(p-j-1)}({\mathcal K}_t)-b_{p-j}\right)
\mbox{.}
\label{zeta}
\end{equation}
\\
The integrals related to the identity orbital integrals can be written as
follows
$$
\int_0^{\infty}dt\,t^{s-1}
I_{\Gamma}^{(p-j)}({\mathcal K}_t)
= \frac{\chi(1){\rm Vol}(\Gamma\backslash G)}{2^{2(n-1)}\Gamma(n/2)^2}
\left(
\begin{array}{c}
n-1\\ p-j
\end{array}
\right) \sum_{\ell=0}^{n/2-1}a_{2\ell}^{(p-j)}
$$
\begin{equation}
\times \int_o^{\infty}dt\,t^{s-1}
e^{-t(\alpha -j)}
\int_{\Bbb R}dr\,r^{2\ell+1}e^{-tr^2}{\rm tanh}(\pi r)
\mbox{,}
\end{equation}
where $\alpha \equiv p+\rho_0^2$. Using the identities
\begin{eqnarray}
1 - {\rm tanh}(\pi r) &=& \frac{2}{1+\exp (2\pi r)}
\mbox{,} \notag \\
\int_0^{\infty}\frac{r^{2\ell-1}dr}{1+e^{2\pi r}} &=&
\frac{(-1)^{\ell-1}}{4\ell}(1-2^{1-2\ell})B_{2\ell}
\mbox{,}
\end{eqnarray}
where $B_\ell$ are the Bernoulli numbers, we obtain:
\begin{equation}
\int_{\Bbb R}dr\,r^{2\ell+1}e^{-tr^2}{\rm tanh}(\pi r)
=\ell !t^{-\ell -1}-\sum_{k=0}^{\infty}\frac{(-1)^{\ell}(1-2^{-2\ell -2k-1})t^k}
{k!(\ell +k+1)}B_{2(\ell +k+1)}
\end{equation}
and

\begin{equation*}
\!\!\!\!\!\!\!\!\!\!\!\!\!\!
\int_0^{\infty}dt\,t^{s-1}
I_{\Gamma}^{(p-j)}({\mathcal K}_t)
=\frac{\chi(1){\rm Vol}(\Gamma\backslash G)}{2^{2(n-1)}\Gamma(n/2)^2}
\left(
\begin{array}{c}
n-1\\ p-j
\end{array}
\right) \sum_{\ell=0}^{n/2-1}a_{2\ell}^{(p-j)}
\end{equation*}
\begin{equation}
\times
\left\{
\frac{\ell !\Gamma(s-\ell -1)}{(\alpha - j)^{s-\ell-1}}
- \sum_{k=0}^{\infty}\frac{(-1)^{\ell}(1-2^{-2\ell -2k-1}\Gamma(k+s))}
{k!(\ell +k+1)}
\frac{B_{2(\ell +k+1)}}{(\alpha - j)^{k+s}}
\right\}
\mbox{.}
\end{equation}
The contribution associated with the identity integral at the
point $s=0$ becomes
$$
\lim_{s \rightarrow 0}\frac{1}{\Gamma(s)}\int_0^{\infty}dt\,t^{s-1}
I_{\Gamma}^{(p-j)}({\mathcal K}_t)
= \frac{\chi(1){\rm Vol}(\Gamma\backslash G)}{2^{2(n-1)}\Gamma(n/2)^2}
\left(
\begin{array}{c}
n-1\\ p-j
\end{array}
\right)
$$
\begin{equation}
\times
\sum_{\ell=0}^{n/2-1}a_{2\ell}^{(p-j-1)}\frac{(-1)^{\ell +1}}{\ell +1}
\left((1-2^{-2\ell -1})B_{2(\ell +1)}
+ [\alpha+j+1]^{\ell+1}\right)
\mbox{.}
\label{Id}
\end{equation}
The hyperbolic orbital integrals can be rewriten in terms of
McDonald functions $K_{\nu}(z)$,
\begin{equation}
K_{\nu}(z) = 2^{-\nu-1}z^{\nu}\int_0^{\infty}dtt^{-\nu-1}
e^{-t-z^2/(4t)},\,\,\,\,\,
|{\rm arg}\,z| <\pi/2\,,\,\,\, \Re\,z^2>0
\mbox{.}
\end{equation}
\\
The result is:

$$
\!\!\!\!\!\!\!\!\!\!\!\!\!\!\!\!\!\!\!\!\!\!\!\!\!\!\!\!\!\!\!\!
\!\!\!\!\!\!\!\!\!\!\!\!\!\!\!\!\!\!\!\!\!\!\!\!\!\!\!\!\!\!\!\!
\!\!\!\!\!\!\!\!\!\!\!\!\!\!\!\!\!\!\!\!\!\!\!\!\!\!\!\!\!\!\!\!
\!\!\!\!\!\!\!\!\!\!\!\!\!\!\!\!\!\!\!\!\!\!\!\!\!\!\!\!\!\!\!\!
\int_{0}^{\infty}dt\,t^{s-1}H_{\Gamma}^{(p-j)}({\mathcal K}_t)
$$
\begin{equation}
\!\!
= \sum_{\gamma\in C_
\Gamma-\{1\}}\frac{\chi(\gamma)}{\sqrt{\pi} j(\gamma)}t_\gamma C(\gamma)
\chi_{\sigma_{p-j}}(m_\gamma)
\left(\frac{2\sqrt{\alpha+j}}{t_{\gamma}}\right)^{-s+1/2}\!\!\!\!\!
K_{-s+1/2}\left(t_{\gamma}\sqrt{\alpha - j}\right)
\mbox{.}
\label{orb}
\end{equation}
\\
The analysis of the integral Eq. (\ref{orb}) gives the the
following result (see also Refs.
\cite{Elizalde,Bytsenko1,Bytsenko3,Bytsenko2}): the terms
associated with the hyperbolic orbital integrals vanish when
$s=0$. Finally, using Eqs. (\ref{tensor1}), (\ref{zeta}) and
(\ref{Id}), we get for the conformal anomaly the following
explicit formula, in terms of the dimension $n$ of the hyperbolic
space, the order $p$ of the form, the radius $R$ of the compact
spatial section, the value of $\alpha$ (which will depend on the
nature of the field, see below), Miatello coefficients, and
Bernoulli numbers:

\begin{eqnarray}
<T_\mu^\mu (x) > &=&
\frac{1}{(4\pi)^{n/2}\Gamma(n/2)R^n}
\sum_{j=0}^p(-1)^j\left\{
\sum_{\ell=0}^{n/2-1}\frac{(-1)^{\ell+1}}{\ell+1}
\left(
\begin{array}{c}
n-1\\ p-j
\end{array}
\right)\right. \notag \\
&\times&
\left.
\left[a_{2\ell}^{(p-j)}\left[(1-2^{-2\ell-1})B_{2(\ell+1)}+
(\alpha - j)^{\ell+1}\right]\right.\right. \notag \\
&&\hspace{-12mm} \left.\left. + a_{2\ell}^{(p-j-1)}\frac{p-j}{n-p}
\left[(1-2^{-2\ell-1})B_{2(\ell+1)}+ (\alpha -j
-1)^{\ell+1}\right]\right]\right\} \mbox{.} \label{e37}
\end{eqnarray}
This constitutes the main result of the present paper.
\bigskip

\noindent{\bf The case of a conformally invariant scalar field}.
Restoring now the dependence on the radius $R$, for the specific
case of a minimally coupled scalar field of mass $m$, we have:
$p=j=0$, $\alpha \Rightarrow \alpha+ R^2m^2$, and $\alpha =
\rho_0^2$. For the case of a conformally invariant scalar field,
we have: $\alpha = \rho_0^2 + (n-2)R^2R(x)/[4(n-1)]$, where
$R(x)=-n(n-1)R^{-2}$ is the scalar curvature. Therefore, the final
result is in this case
\begin{eqnarray}
<T_\mu^\mu (x)> &=& \frac{1}{(4\pi)^{n/2}\Gamma (n/2) R^n}
\sum_{\ell=0}^{n/2-1} \frac{(-1)^{\ell+1}}{\ell+1}a_{2\ell}
\notag \\
&\times& \left[ 2^{-2\ell-2} + (1-2^{-2\ell-1}) B_{2\ell+2}
\right] \mbox{.}
\end{eqnarray}
This formula is in full agreement with a previous  result obtained
in Ref. \cite{Bytsenko1} and constitutes a check of our main
formula Eq. (\ref{e37}). In fact, we obtain from this expression
Table 1 [note a small missprint in the denominator of the last
value given in the table in Ref. \cite{Bytsenko1}].

\begin{table}

\begin{center}

\begin{tabular}{|c||c|c|} \hline \hline
  $<T_\mu^\mu (x)>_{\mbox{c.i.s.}}$  & exact & numerical \\ \hline
\hline       &                                 & \\
$n=2$ & $-\displaystyle\frac{1}{12\, \pi}$ & $ - 0.0265258 $ \\
 &                                 & \\
 \hline       &                                 & \\
 $n=4$ & $-\displaystyle\frac{1}{240\, \pi^2}$ & $ - 4.22172 \times 10^{-4} $ \\
 &                                 & \\
 \hline       &                                 & \\
  $n=6$ & $-\displaystyle\frac{5}{4032\, \pi^3}$ & $ - 3.99945 \times 10^{-5} $ \\
 &                                 & \\
 \hline       &                                 & \\
  $n=8$ & $-\displaystyle\frac{23}{34560\, \pi^4}$ & $ -6.83210 \times 10^{-6} $ \\
 &                                 & \\
 \hline       &                                 & \\
 $n=10$ & $-\displaystyle\frac{263}{506880\, \pi^5}$ & $-1.69551 \times 10^{-6}$ \\
 &                                 & \\
 \hline       &                                 & \\
$n=12$ & $-\displaystyle\frac{133787}{251596800\, \pi^6}$ & $-5.53107 \times 10^{-7}$ \\
 &                                 & \\
 \hline       &                                 & \\
 $n=14$ & $-\displaystyle\frac{157009}{232243200\, \pi^7}$ &
       $-2.23837 \times 10^{-7} $ \\
 &                                 & \\
\hline \hline \end{tabular} \bigskip

\caption{{\protect\small Exact and numerical values of the
conformal anomaly for the conformally invariant scalar field, in
dimensions $n=2$ to $n=14$ (we have set $R=1$).}}

\end{center}

\end{table}

\medskip

\noindent {\bf Explicit and numerical values of the conformal
anomaly for $p-$forms}. Using our Eq.~(\ref{e37}), exact explicit
values and also numerical values of the conformal anomaly
corresponding to spaces of arbitrary dimension $n$ and forms of
any order $p$ are easily obtained, with the help of any standard
program as Matlab, Maple or Mathematica. Using Mathematica 5.0 on
a laptop, in a question of seconds we have obtained the following
table (Table 2) for the conformal anomaly, where we have set $R=1$
and $\alpha = p + \rho_0^2$, with $\rho_0=(n-1)/2$.

\begin{table}

\begin{center}

\hspace*{-12mm}\begin{tabular}{|c||c|c|c|c|c|} \hline \hline
  $<T_\mu^\mu (x)>$  & $p=0$ & $p=1$ & $p=2$ & $p=3$ & $p=4$ \\ \hline
\hline       &                                 & & & & \\
       $n=2$ & $-\displaystyle\frac{1}{12\, \pi}$ & & & & \\
             &                                 & & & & \\
             & = - 0.0265258                  & & & & \\
             &                                 & & & & \\
\hline       &                                 & & & & \\
       $n=4$ & $\displaystyle\frac{29}{240\, \pi^2}$
& $-\displaystyle\frac{67}{160\, \pi^2}$ & & & \\
             &                                 & & & & \\
             & = 0.012243                      & = - 0.0424282 & & & \\
             &                                 & & & & \\
\hline       &                                 & & & & \\
       $n=6$ & $-\displaystyle\frac{1139}{4032\, \pi^3}$
& $\displaystyle\frac{2539}{2016\, \pi^3}$
& $-\displaystyle\frac{2005}{1792\, \pi^3}$ & & \\
             &                                 & & & & \\
             & = - 0.00911074                  & = 0.0406184
             & = - 0.036085                    & & \\
             &                                 & & & & \\
\hline       &                                 & & & & \\
$n=8$ & $\displaystyle\frac{32377}{34560\, \pi^4}$ &
$-\displaystyle\frac{1368853}{276480\, \pi^4}$ &
 $\displaystyle\frac{101665}{41472\, \pi^4}$
& $\displaystyle\frac{118459}{34560\, \pi^4}$ & \\
             &                                 & & & & \\
             & = 0.00961753                    & = - 0.0508269
             & = 0.0251662                     & = 0.035188 & \\
             &                                 & & & & \\
\hline       &                                 & & & & \\
$n=10$ & $-\displaystyle\frac{2046263}{506880\, \pi^5}$ &
$\displaystyle\frac{16454263}{675840\, \pi^5}$ &
 $-\displaystyle\frac{2475365}{811008\, \pi^5}$
& $-\displaystyle\frac{34196177}{7096320\, \pi^5}$
&$-\displaystyle\frac{14020681}{135168\, \pi^5}$  \\
             &                                 & & & & \\
             & = - 0.0131919                   & = 0.0795582
             & = - 0.00997389                  & = - 0.0157469
             & = - 0.338958 \\
             &                                 & & & & \\
\hline \hline \end{tabular}
\medskip

\caption{{\protect\small Exact and numerical values of the conformal
anomaly for the family of spacetimes of dimension $n=2$ to $n=10$
which possess a compact spatial section, corresponding to forms of
order up to $p=4$.}}

\end{center}

\end{table}

\section{Conclusions}

We have here evaluated the conformal anomaly for the family of
spacetimes of arbitrary dimension which possess a compact spatial
section of the form $\Gamma\backslash {\Bbb H}^n$. We have
restricted ourselves to the situation where the manifold is smooth
and $\Gamma$ is a discrete subgroup of $SO_1(n,1)$, acting freely
and properly discontinuously on ${\Bbb H}^n$. The terms associated
with hyperbolic orbital integrals do not contribute to the
conformal anomaly, as we have shown above.

Explicit exact and numerical results for the conformal anomaly
corresponding to $p-$forms of orders $p=0$ to $p=4$ in spaces of
dimension $n=2$ to $n=10$ have been given in Table 2. Both the
sign and the magnitude of the anomaly seem to change in a rather
non--uniform way in the cases considered, their absolute value
being always less than 1 for the calculated cases (but this can be
shown to be not a bound for forms of higher order). In fact, one
sees clearly, that the absolute value of the conformal anomaly for
$p-$forms definitely increases with the order of the form in
hyperbolic spaces of higher dimensionality, of the class
considered.

As a particular case, we recover the formula for the conformal
invariant scalar field in any dimension and, in a similar way, a
number of more general situations can be treated with the same
techniques as described in this paper.

\end{document}